\documentclass[12pt,preprint]{aastex62}
\usepackage{rotating}
\usepackage{epsfig}

\begin{document}

\title{Proper Motions of the Radio Source Orion MR, Formerly Known as Orion n, and New Sources with Large Proper Motions in Orion BN/KL}

\author{Luis F. Rodr{\'\i}guez}
\affil{Instituto de Radioastronom\'\i a y Astrof\'\i sica, 
Universidad Nacional Aut\'onoma de M\'exico, Apdo. Postal 3-72 (Xangari), 58089 Morelia, Michoac\'an, M\'exico}
\affil{Mesoamerican Center for Theoretical Physics, Universidad
Aut\'onoma de Chiapas, Carretera Emiliano Zapata km. 4,
Real del Bosque (Ter\'an). 29050 Tuxtla Guti\'errez, Chiapas,
M\'exico}

\author{Sergio A. Dzib}
\affil{Max-Planck-Institut f\"ur Radioastronomie, Auf dem H\"ugel 69, D-53121 Bonn, Germany}

\author{Luis Zapata}
\affil{Instituto de Radioastronom\'\i a y Astrof\'\i sica, 
Universidad Nacional Aut\'onoma de M\'exico,, Apdo. Postal 3-72 (Xangari), 58089 Morelia, Michoac\'an, M\'exico}

\author{Susana Lizano}
\affil{Instituto de Radioastronom\'\i a y Astrof\'\i sica, 
Universidad Nacional Aut\'onoma de M\'exico,, Apdo. Postal 3-72 (Xangari), 58089 Morelia, Michoac\'an, M\'exico}

\author{Laurent Loinard}
\affil{Instituto de Radioastronom\'\i a y Astrof\'\i sica, 
Universidad Nacional Aut\'onoma de M\'exico,, Apdo. Postal 3-72 (Xangari), 58089 Morelia, Michoac\'an, M\'exico}
\affil{Instituto de Astronom\'\i a, Universidad Nacional Aut\'onoma de M\'exico, Apartado Postal 70-264, CdMx, CP 04510, M\'exico}

\author{Karl M. Menten}
\affil{Max-Planck-Institut f\"ur Radioastronomie, Auf dem H\"ugel 69, D-53121 Bonn, Germany}

\author{Laura G\'omez}
\affil{Joint ALMA Observatory, Alonso de C\'ordova 3107, Vitacura, Santiago, Chile}

\email{l.rodriguez@irya.unam.mx}
 
\begin{abstract}

The infrared source known as Orion n was detected in 1980 with observations made with the 3.8-m United Kingdom Infrared Telescope.
About two decades later, sensitive observations made with the Very Large Array revealed the presence of a mJy double radio source apparently
coincident in position with the infrared source n. The radio source was assumed to be the counterpart of the infrared source. However, over the years
it has been concluded that the radio source shows large proper motions to the south while the infrared source n is stationary.
Here we reanalyze the proper motions of the radio source adding both older and newer VLA observations than previously used.
We confirm the proper motions of the radio source that at present no longer coincides positionally with the infrared source. 
The solution to this problem is, most probably, that the infrared source n and the
radio source are not the same object: the infrared source is a stationary object in the region while the radio counterpart is moving as
a result of the explosion that took place in this region some 500 years ago and that expelled large amounts of molecular gas as well
as several compact sources. Considering the paper where it was first reported, we refer to this double radio source as Orion MR.

In addition, we use these new observations to fully confirm the large proper motions of the sources IRc23 and Zapata 11. Together with sources BN, I, Orion MR, and x, there are at least six compact sources that
recede from a point in common in Orion BN/KL. However, IRc23 is peculiar in that its ejection age appears to be only $\sim$300 years.
The relatively large number of sources rules out as a possible mechanism the classic three-body 
scenario since then only two escaping bodies are expected: a tight binary plus the third star involved in the encounter.

\end{abstract}  

\keywords{Stars: individual (Orion n) --
stars: radio continuum 
}

%\end{document}
 
\section{Introduction}

The Orion BN/KL region hosts a remarkable explosive phenomenon whose nature is yet unclear.
Both molecular gas (Bally et al. 2017; Youngblood et al. 2018) and compact objects (Rodr\'\i guez et al. 2017; Luhman et al. 2017) appear to have been ejected some 500 years ago from a common central point.
The compact objects that have well measured proper motions are BN, I, n, and more recently source x, studied in the near-infrared by Luhman et al. (2017).
In addition, Dzib et al (2017) proposed that the sources Zapata 11 and IRc 23 may also be part of the group of receding sources.

One of the sources that has received particular attention is the source n. While the proposed radio counterpart seems to have well determined proper motions (Rodr\'\i guez et al. 2017),
recent observations suggest that the infrared source is stationary (Luhman et al. 2017; Kim et al. 2019; Pagani et al. 2019) and it is unlikely to have participated in the explosive event.

In this paper we present four sets of observations in addition to those discussed by Rodr\'\i guez et al. (2017). These authors used VLA observations from 1991.67 to 2014.17 to discuss the
proper motions of the radio source apparently associated to source n. Here we add an older observation (1985.19) and three more recent ones (2016.91, 2016.95 and 2018.38) to readdress the issue.
We use these observations also to further study the proposed proper motions of Zapata 11 and IRc23.
%We then improved the time coverage, $\Delta t$,  from 22.5 years to 33.2 years. 
%Since the error in the proper motions decreases as $(\Delta t)^{-3/2}$ (Dzib et al 2017) we obtain a significant improvement of
%order 2 in the signal-to-noise of our results.

\section{Observations}

In Table 1 we present the parameters of the new four projects taken into account in this study. The first three were taken from the VLA archive, while the latter one was part of our
VLA project 17A-069, made with the Karl G. Jansky Very Large Array (VLA) of NRAO\footnote{The National 
Radio Astronomy Observatory is a facility of the National Science Foundation operated
under cooperative agreement by Associated Universities, Inc.}. All observations were made in the A configuration at the epochs and frequencies
given in Table 1.  The observations of projects AG177 and 16B-233 were made in seven and five epochs nearby in time, respectively,  that were concatenated to
produce a single image. These images were assumed to have an epoch equal to the mean of all epochs considered in each project.
The flux and bandpass calibrator was J1331+305 or J0137+331 and
the phase calibrator was J0541$-$054.

The 1985.19 observations were obtained with a total bandwidth of 100 MHz and reduced using the software AIPS (Astronomical Image Processing System).
The position of J0541$-$054 was corrected to the latest updated position in the VLA calibrator manual to ensure accurate
astrometry.
The three recent data sets were obtained and analyzed as follows.
The digital correlator of the VLA was configured in spectral windows of 128 MHz width, each divided 
in 64 channels of spectral resolution of 2 MHz. The total bandwidth for each data set is given in Table 1.
The data were analyzed in the standard manner using the CASA (Common Astronomy Software Applications) package of NRAO using
the pipeline provided for VLA\footnote{https://science.nrao.edu/facilities/vla/data-processing/pipeline} observations. 
Maps were made using a robust weighting (Briggs 1995) of 0 in order to optimize the compromise between sensitivity and angular resolution. 
All images were made using only baselines larger that 75 k$\lambda$, to suppress the extended emission of the region for scales larger than
$\sim 3''$.
The angular resolution of the observations is given also in Table 1.

\clearpage

\begin{deluxetable}{lccccc}
\tabletypesize{\scriptsize}
% \tablewidth{18.5cm}
\tablecaption{New VLA and Jansky VLA Data Used for the Determination of the Proper Motions
of the Radio Source\tablenotemark{a}}
%\small
\tablehead{
\colhead{}  & \colhead{Mean}  & \colhead{} & \colhead{Frequency} 
& \colhead{Bandwidth} & \colhead{Synthesized Beam} \\
\colhead{Epochs}  &\colhead{Epoch} & \colhead{Project} & \colhead{(GHz)}
& \colhead{(GHz)} & \colhead{($\theta_M \times \theta_m; PA$)\tablenotemark{b}}}
\startdata
1985 Feb 16, Mar 01, 06,  08, 15, 18, April 01 & 1985.19 & AG177 & 4.86 & 0.1 & $0\rlap.{''}45\times0\rlap.{''}42;-68^\circ$ \\
2016 Nov 27 & 2016.91 & 16B-268 & 6.10 & 2.0 & $0\rlap.{''}30\times0\rlap.{''}22;-27^\circ$ \\
2016 Nov 21, Dec 02, 06,  23, 2017 Jan 04 & 2016.95 & 16B-233 & 5.50 & 2.0 & $0\rlap.{''}35\times0\rlap.{''}29;+46^\circ$ \\
2018 May 11 & 2018.38 & 17A-069 & 6.00  & 4.0 & $0\rlap.{''}33\times0\rlap.{''}24;-28^\circ$ \\
\enddata
\tablenotetext{a}{All observations were made in the A configuration.}
\tablenotetext{b}{Major axis$\times$minor axis in arcsec; PA in degrees.}
%% You can append references to a table using the \tablerefs command.
\end{deluxetable}

%\clearpage

\begin{deluxetable}{lcccc}
\tabletypesize{\scriptsize}
% \tablewidth{18.5cm}
\tablecaption{Parameters
of the Radio Source Orion MR}
%\small
\tablehead{
\colhead{}  & \multicolumn{2}{c}{Position$^a$} 
& \colhead{Flux Density} & \colhead{Deconvolved Size} \\
\colhead{Epoch} & \colhead{$\alpha(2000)$} & \colhead{$\delta(2000)$}
& \colhead{(mJy)} & \colhead{($\theta_M \times \theta_m; PA$)\tablenotemark{b}}}
\startdata
1985.19 &  $14\rlap.{^s}356\pm0\rlap.{^s}001$ &  $32\rlap.{''}63\pm0\rlap.{''}02$ & 0.63$\pm$0.11& $0\rlap.{''}25\times0\rlap.{''}00;-4^\circ\pm4^\circ$ \\
2016.91 &  $14\rlap.{^s}354\pm0\rlap.{^s}001$ &  $32\rlap.{''}84\pm0\rlap.{''}01$& 1.54$\pm$0.10 & $0\rlap.{''}52\times0\rlap.{''}00;+31^\circ\pm3^\circ$ \\
2016.95 & $14\rlap.{^s}359\pm0\rlap.{^s}001$ &  $32\rlap.{''}80\pm0\rlap.{''}01$& 1.83$\pm$0.16 & $0\rlap.{''}44\times0\rlap.{''}00;+30^\circ\pm6^\circ$ \\
2018.38 &  $14\rlap.{^s}354\pm0\rlap.{^s}001$ &  $32\rlap.{''}85\pm0\rlap.{''}01$& 1.02$\pm$0.10 & $0\rlap.{''}55\times0\rlap.{''}00;+29^\circ\pm4^\circ$ \\
\enddata
\tablenotetext{a}{$\alpha(2000) = 05^h~35^m$; $\delta(2000) = -05^\circ~22'$.}
\tablenotetext{b}{Major axis$\times$minor axis in arcsec; PA in degrees. The deconvolved dimensions of the source are corrected for the effect of bandwidth smearing.
In all cases the minor axis of the source remains unresolved.}
%% You can append references to a table using the \tablerefs command.
\end{deluxetable}

\begin{figure}
\centering
\vspace{-2.8cm}
\includegraphics[angle=0,scale=0.5]{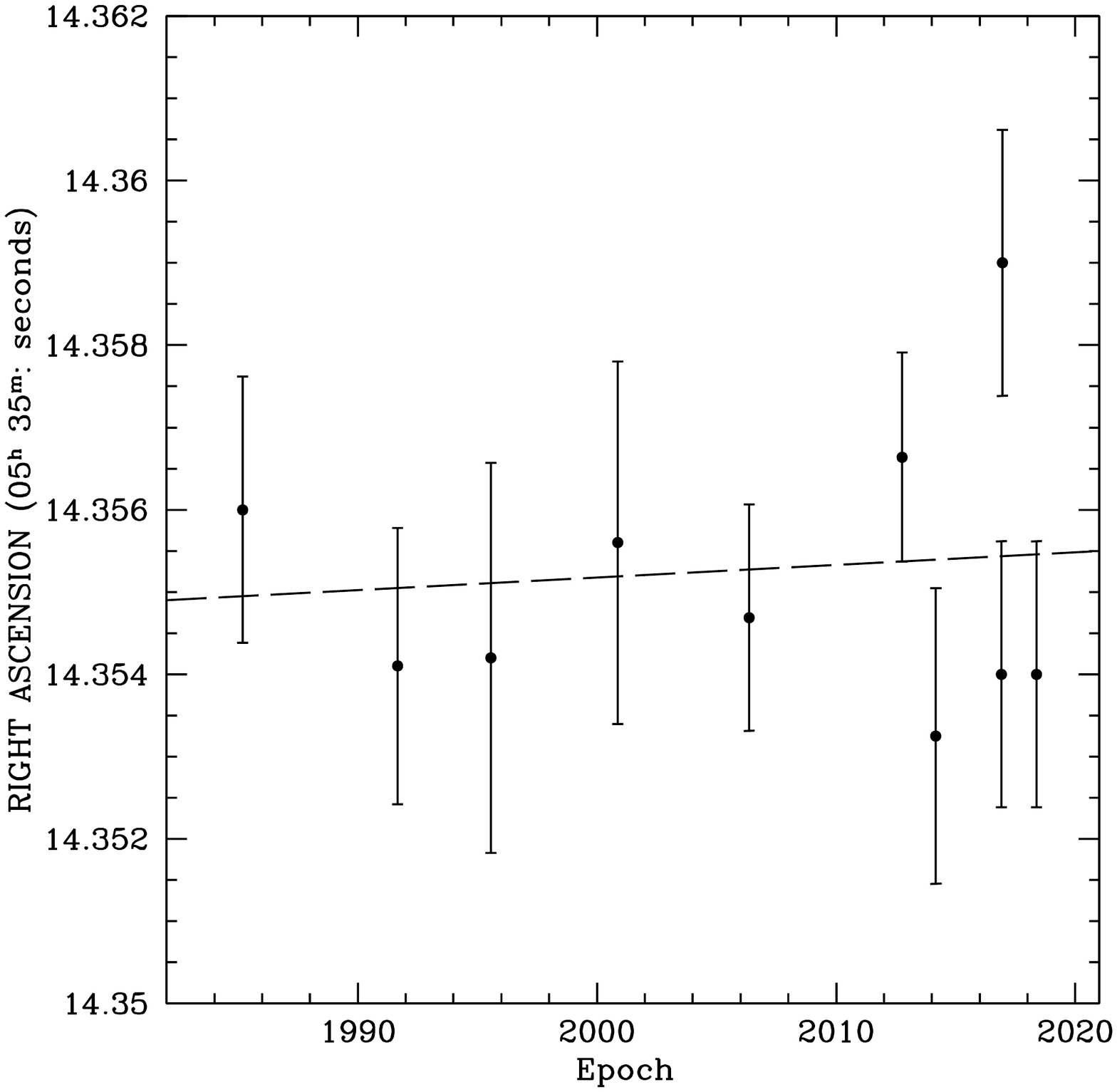}
\vskip-4.0cm
\includegraphics[angle=0,scale=0.5]{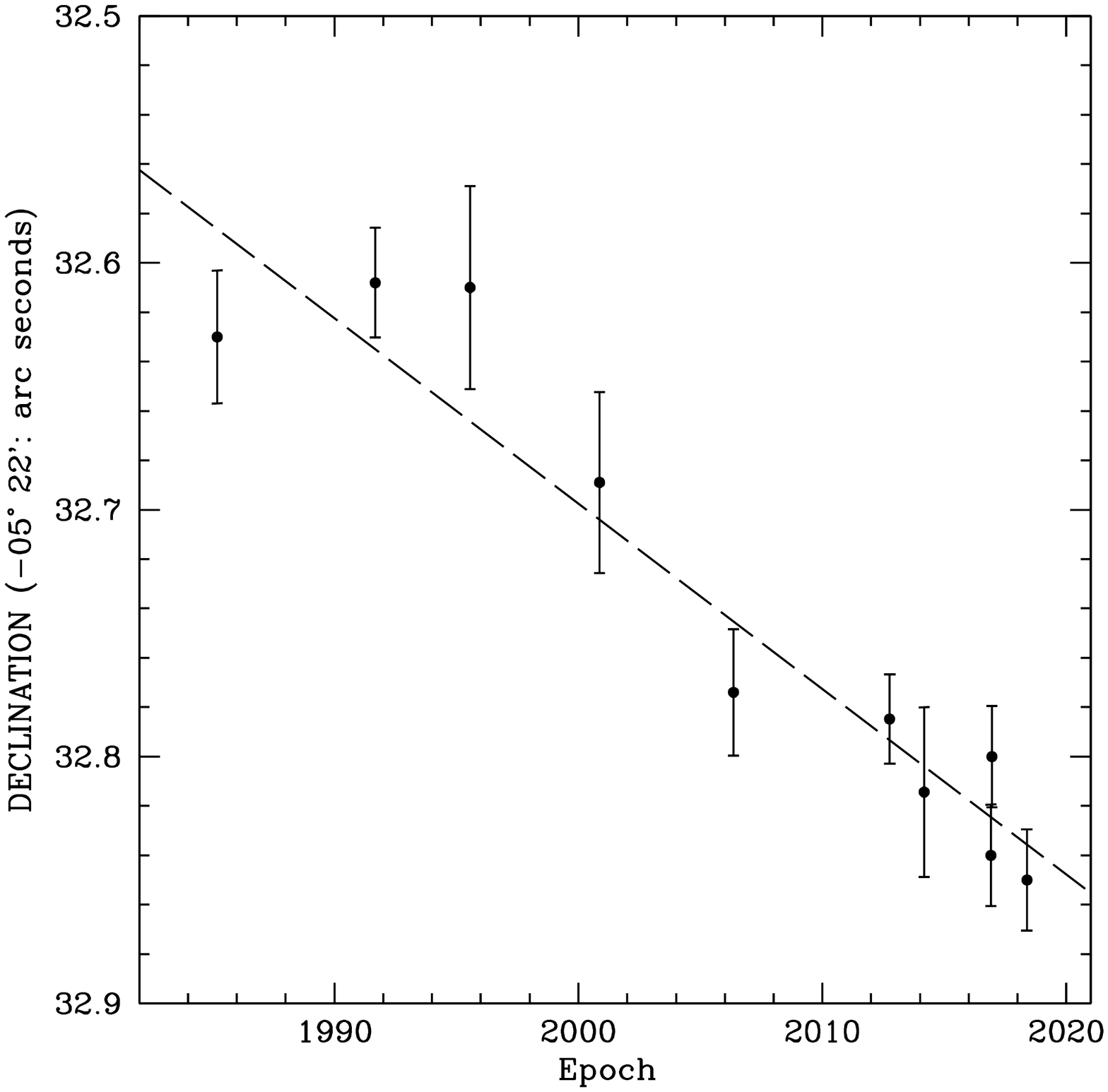}
\vskip-1.5cm
\caption{\small Right ascension (top) and declination (bottom) of the source Orion MR as a function of time. The error bars include a systematic component
added in quadrature to make $\chi^2$ = 1. The dashed lines are least-squares linear fits to the data. The parameters of the fits are given in the text. The proper motion in
right ascension is consistent with a null result while the proper motion in declination is detected at a 10-$\sigma$ level of significance.
}
\label{fig1}
\end{figure}
%\pagebreak

\clearpage

\section{Proper Motions of the Radio Source Orion MR}

The radio source was fitted with a single Gaussian ellipsoid and the positions obtained at different epochs are listed in Table 2. 
In Figure 1 we show the right ascension and declination of the source as a function of time, using both the 6 epochs reported by Rodr\'\i guez et al. (2017)
and the 4 new epochs presented here. In the case of epoch 1991.67, where two peaks are present, we used the average value of the position of the two peaks.
A least-squares linear fit to the positions was made, adding in quadrature a systematic noise to the statistical noise of the positions
in order to obtain a $\chi^2$ value of 1 in the fit. This systematic contribution was $0\rlap.^s00127$ in right ascension and $0\rlap.{''}018$ in declination.

The position for epoch 2000 from this fit is:

$$\alpha(2000) = 05^h~ 35^m~ 14\rlap.^s3552 \pm 00\rlap.^s0006; $$
$$\delta(2000) = -05^\circ~ 22'~ 32\rlap.{''}698 \pm 00\rlap.{''}010,$$

\noindent while the proper motions are:

$$\mu_\alpha cos~ \delta = 1.4 \times 10^{-2} \pm 4.7 \times 10^{-2}~msec~yr^{-1};$$
$$\mu_{\delta} =  -7.5 \pm 0.7~mas~yr^{-1}.$$

From these last two determinations we conclude that while the proper motion in right ascension is consistent with a
null result, the proper motion in declination is robustly detected, at a 10-$\sigma$ level of significance. At a distance of 388 pc (Kounkel et al. 2017),
the declination proper motion corresponds to a velocity of 14$\pm$2 km s$^{-1}$ in the plane of the sky. These values are
consistent at the $\pm$1-$\sigma$ level with those reported by Rodr\'\i guez et al. (2017). Of course, this is not surprising since
there are six epochs in common in the studies.

Rodr\'\i guez et al. (2017) proposed that the explosion that ejected the stars took place around the year 1475 at position:

$$\alpha(2000) = 05^h~ 35^m~ 14\rlap.^s41; $$
$$\delta(2000) = -05^\circ~ 22'~ 28\rlap.{''}2.$$

If we displace to year 1475 the 2000 epoch position of the radio source using the proper motions given above, we obtain:

$$\alpha(1475) = 05^h~ 35^m~ 14\rlap.^s36; $$
$$\delta(1475) = -05^\circ~ 22'~ 28\rlap.{''}8.$$

Since both positions coincide within $\sim$1$''$, we conclude that the hypothesis that the radio source Orion MR was ejected simultaneously
with the other compact sources is tenable.

\begin{figure}
\centering
\vspace{-2.8cm}
\includegraphics[angle=0,scale=0.5]{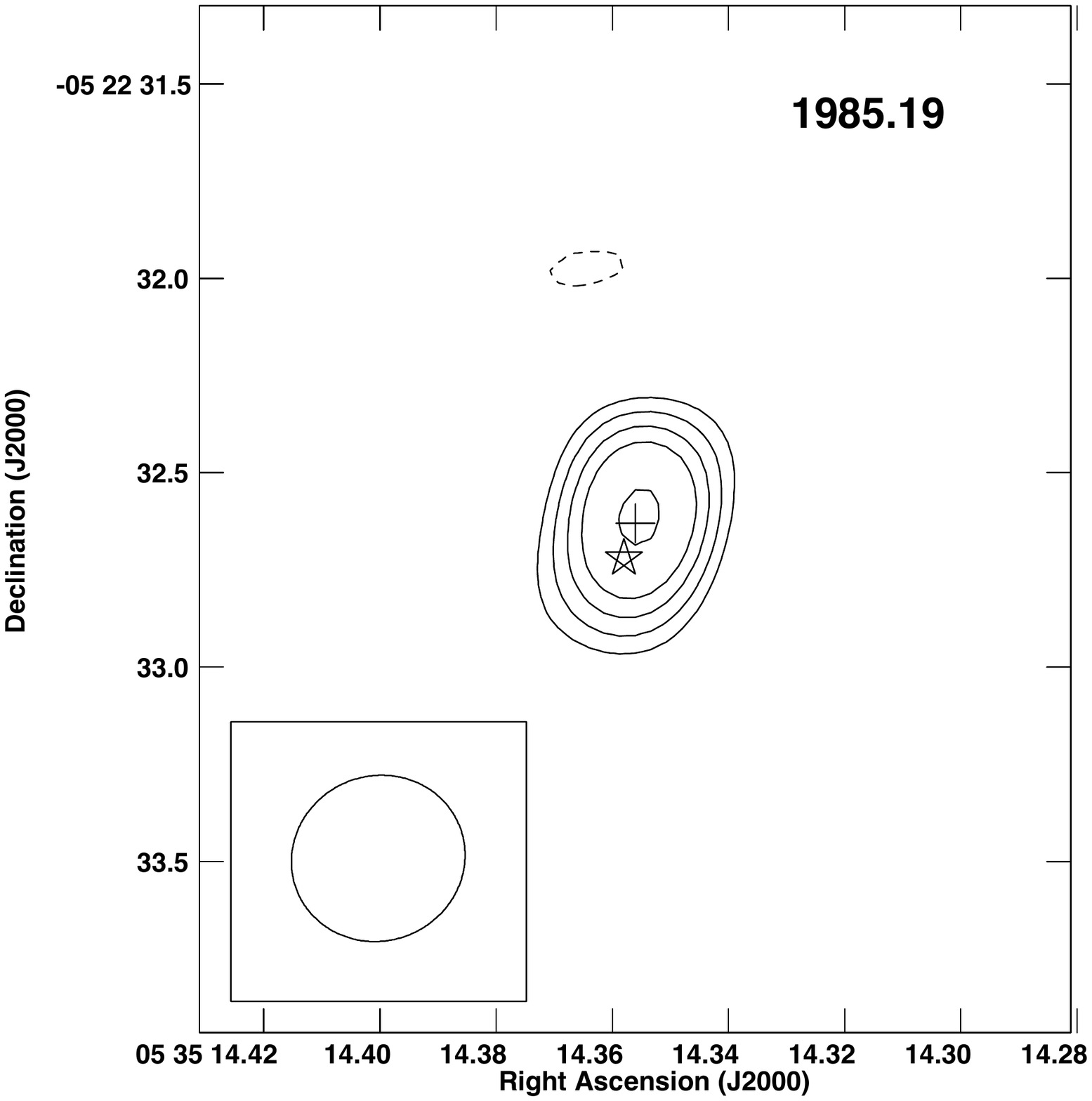}
\vskip-4.4cm
\includegraphics[angle=0,scale=0.5]{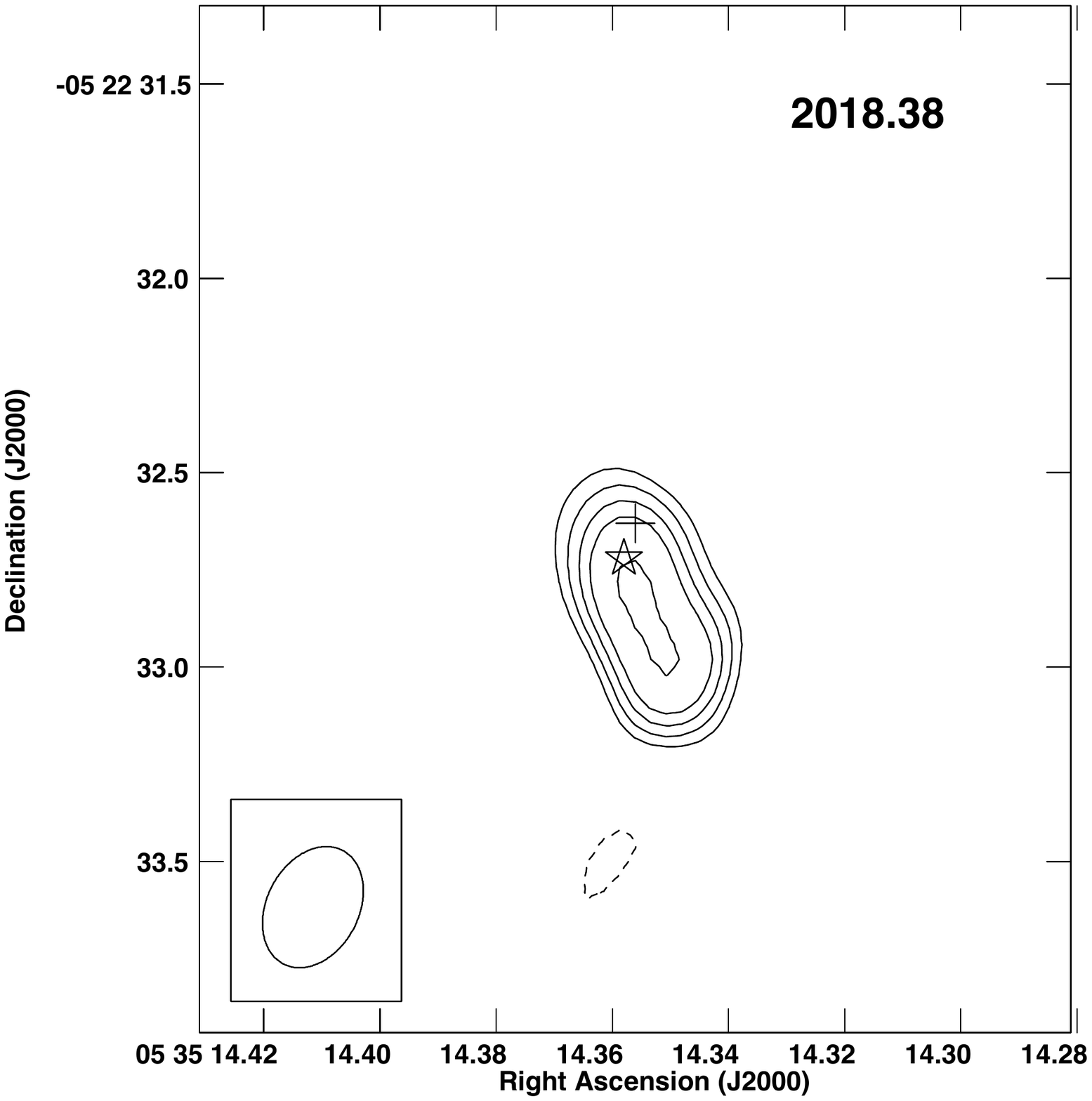}
\vskip-1.5cm
\caption{\small VLA images of Orion MR for epochs 1985.19 (top) and 2018.38 (bottom). The cross marks the peak position of the radio source for 1985.19 and the star
the position of the infrared source n from Kim et al. (2019; the source 703 in their Table 3). Note the displacement to the south of the radio source as well as the change in the position angle of its major axis.
Contours are -4, -3, 3, 4, 5, 6 and 8 times 70 and 40 $\mu$Jy beam$^{-1}$, the rms noise of the 1985.19 and 2018.38 epochs, respectively.
The synthesized beams are shown in the bottom left corner of the images.
}
\label{fig2}
\end{figure}
%\pagebreak

\clearpage

In Figure 2 we show the images of the radio source at the extreme epochs of our coverage: 1985.19 and 2018.38. From this image it is evident that 
the radio source has moved to the south and that at present its peak of emission is no longer coincident with the infrared source n.
We then propose that the most reasonable explanation for the discrepancy between the radio and IR proper motions is
that we are dealing with two different sources. Another evidence in favor of this explanation is that the infrared source n has always appeared
as a single source, while the radio source is clearly seen as double or elongated in images such as those presented by
G\'omez et al. (2005) and Rodr\'\i guez et al. (2017). Since, to our knowledge, the radio source was first reported by Menten \& Reid (1995), we will refer to it as Orion MR.

Another plausible explanation for the discrepancy between the proper motions of source n and Orion MR would be to 
interpret source n as a reflection nebula illuminated by one or both of the radio components of Orion MR. In such an arrangement, a large 
displacement of the illuminating source might not produce a significant change in the extended reflected light. However, 
Simpson et al. (2006) found that the linear polarization of source n at 2 $\mu$m is only 2\%, similar to that of other stars in the field and consistent with
dichroic absorption in the line-of-sight. In contrast, extended sources in the field have linear polarizations in the range of 5 to 75\% and require of local
scattering of radiation.

\begin{deluxetable}{lccc}
\tabletypesize{\scriptsize}
% \tablewidth{18.5cm}
\tablecaption{Deconvolved Position Angle of the Major Axis
of the Radio Source Orion MR}
%\small
\tablehead{
\colhead{} & \colhead{} & \colhead{Frequency}
& \colhead{Position Angle} \\
 \colhead{Epoch} & \colhead{Project} & \colhead{(GHz)}
& \colhead{(Degrees)}}
\startdata
1985.19 &  AG177 & 4.9 & -4$\pm$4\\
1991.67  &  AM335  &  8.4  &  14$\pm$5 \\
1995.56  &  AM494  &  8.4  & 23$\pm$3 \\
2000.87  &  AM688 & 8.5  &  20$\pm$3 \\
2006.36  & AR593  &  8.5  & 20$\pm$3 \\
2012.76 &  SD0630  &  6.1  & 22$\pm$2 \\
2014.17 &  13B-085 &  7.2 & 26$\pm$5 \\
2016.91 &  16B-268  & 6.1 & 31$\pm$3 \\
2016.95 & 16B-233 & 5.5 & 30$\pm$6 \\
2018.38 & 17A-069 & 6.0 & 29$\pm$4 \\
\enddata
%\tablenotetext{a}{$\alpha(2000) = 05^h~35^m$; $\delta(2000) = -05^\circ~22'$.}
%\tablenotetext{b}{Major axis$\times$minor axis in arcsec; PA in degrees. The deconvolved dimensions of the source are corrected for the effect of bandwidth smearing.}
%% You can append references to a table using the \tablerefs command.
\end{deluxetable}

\section{High-angular resolution images of Orion MR}

In most of the images analyzed by us, Orion MR appears as a single elongated source that we have fitted with one Gaussian ellipsoid.
It is only in two of the epochs (1991.67, project AM335 and 2016.95, project 16B-233) that we see clearly the object as a double source.
These images are shown in Figure 3. In both epochs the double source shows a separation of $\sim 0\rlap.{''}3$ and the change in the position angle that is
discussed below in detail is evident. Analysis of Figure 3 suggests a possible alternative interpretation. The source n could be the southern component of
the 1991.67 image and the northern component in the 2016.95 image. In both epochs the second component could be an ejecta, to the south in
1991.67 and to the north in 2016.95. However, from Figure 3 we can see that the southern component of the 1991.67 image and the northern
component of the 2016.95 image are separated by $\sim0\rlap.{''}1$. Is this separation consistent with the proper motions reported for the infrared source n?
The weighted mean proper motions from the results of Luhman et al. (2017), Goddi et al. (2011) and Kim et al. (2019) are 

$$\mu_\alpha cos~ \delta = 0.9 \pm 1.6~mas~yr^{-1};$$
$$\mu_{\delta} =  0.7 \pm 1.6~mas~yr^{-1},$$

\noindent consistent with a stationary source. Over a time span of 25.28 years these proper motions could provide a displacement of order $\sim0\rlap.{''}03$,
a factor of three smaller than the displacement derived from Figure 3.

\begin{figure}
\centering
\vspace{-2.8cm}
\includegraphics[angle=0,scale=0.6]{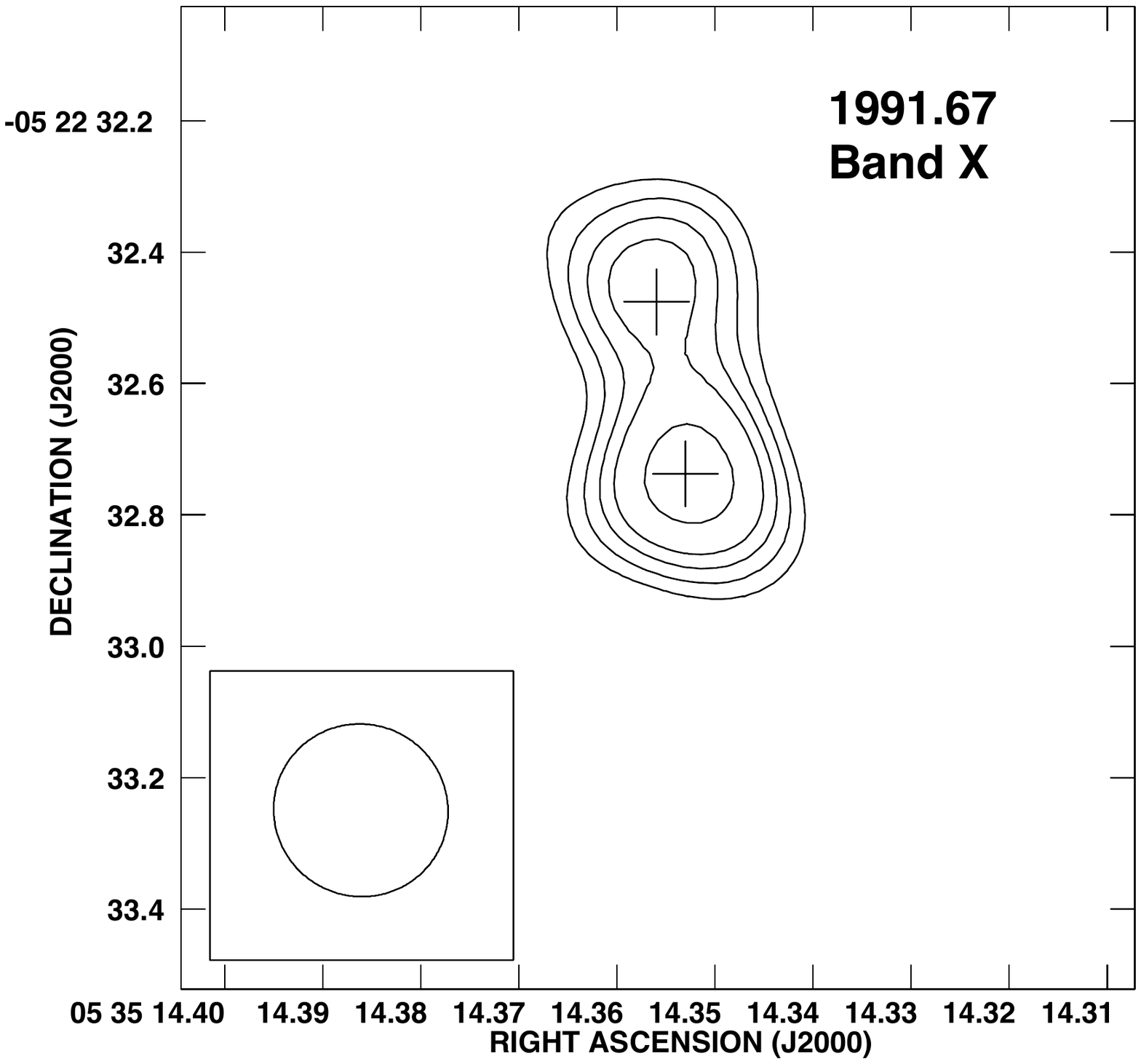}
\vskip-6.6cm
\includegraphics[angle=0,scale=0.6]{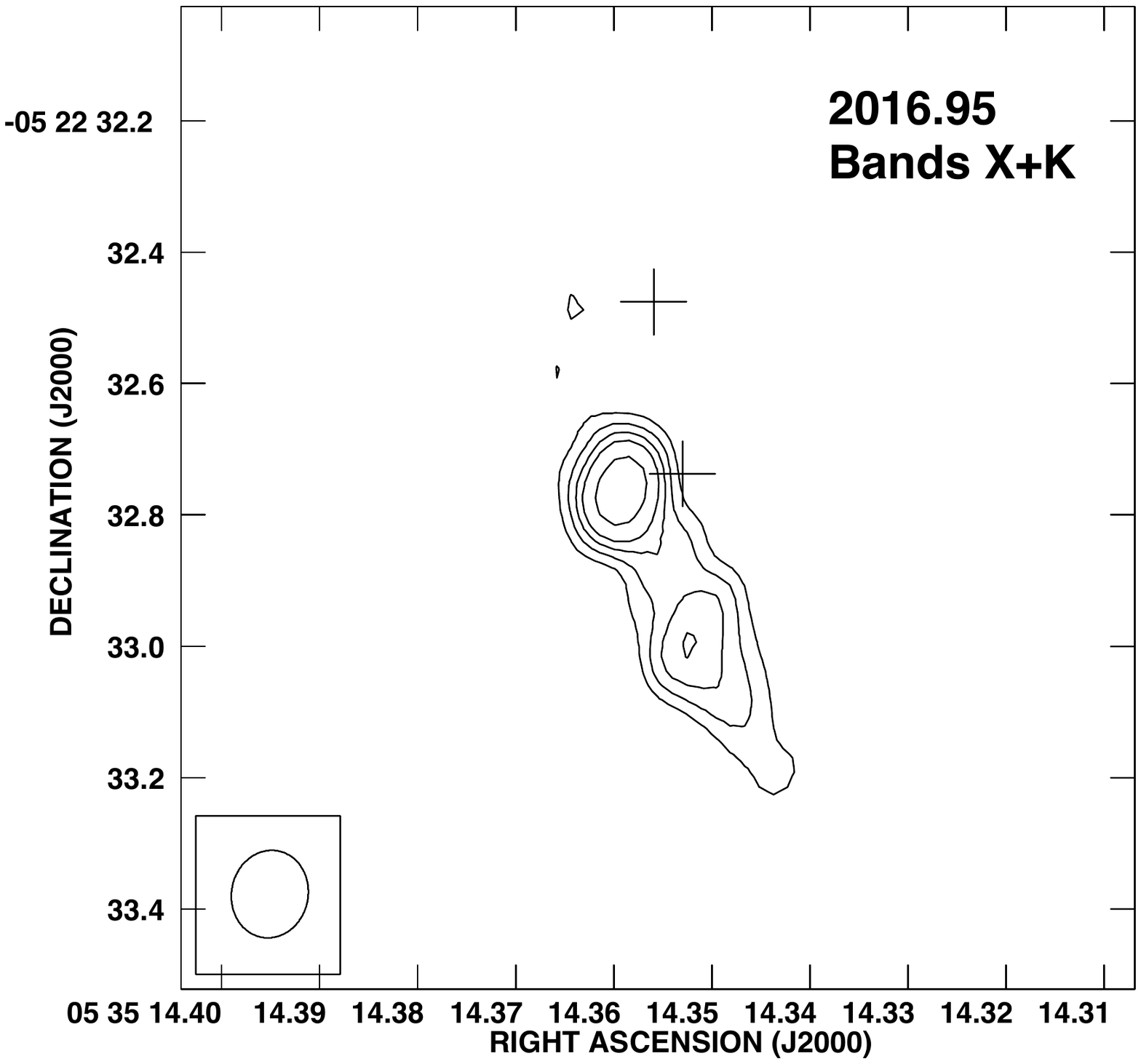}
\vskip-3.0cm
\caption{\small VLA images of Orion MR for epochs 1991.67 (top) and 2016.95 (bottom). The two crosses mark the peak positions of the two components of the radio source for 1991.67.
Note the displacement to the south of the radio source as well as the change in the position angle of its major axis.
Contours are -3, 3, 4, 5, 6 and 8 times 110 and 33 $\mu$Jy beam$^{-1}$, the rms noise of the 1991.67 and 2016.95 epochs, respectively.
The synthesized beams are shown in the bottom left corner of the images.
}
\label{fig3}
\end{figure}

\section{How likely is the fortuitous spatial coincidence of a radio source with a near-IR source?}

Hillenbrand \& Carpenter (2000) detected a total of 778 near-infrared (H and K) stars in the inner $5\rlap.'1 \times 5\rlap.'1$ region of the ONC.
The solid angle of the sampled region is $9.4 \times 10^4$ arcsec$^2$. If we assume that an association, real or apparent, with one of the near-infrared sources
is taken to be for the case that the radio source falls within 1 arcsec of the near-IR position, we find that the solid angle for an association will be
$778 \times \pi$ arcsec$^2$ = $2.4 \times 10^3$ arcsec$^2$. Then the probability that a given position falls within 1 arcsec of a near-IR source will
be the ratio given by ($2.4 \times 10^3$ arcsec$^2$/$9.4 \times 10^4$ arcsec$^2$) = 0.026, or one part in 39. Finally, in a solid angle similar to that
sampled by Hillenbrand \& Carpenter (2000), Forbrich et al. (2016) report a total of 556 compact centimeter sources. We then expect in the order of $556 \times 0.026 \simeq$ 14
associations due just to the high density of near-IR and radio sources. We conclude that the hypothesis of a fortuitous association in the plane of the sky between source n and Orion MR
is not unlikely.

\section{Rotation in the Radio Source?}

As can be seen in Figure 2, the position angle of the major axis of Orion MR seems to have changed by $\sim$30$^\circ$ over the time interval of 33.19 years.
This change in position angle is also evident in Figure 3.
To verify the possible rotation of the position angle, we have plotted the ten values given in Table 3 as a function
of time. A linear fit to the data (including a systematic error of 3.3$^\circ$ for the position angles) is shown in Figure 4. The slope of the fit is 0.73$\pm$0.15 degrees yr$^{-1}$,
indicating that the change is significant at the 4.9-$\sigma$ level. Assuming that this rotation will continue in a constant way, we find that it will be completed with
a period of 490$\pm$100 years. As noted above, G\'omez et al. (2005) have resolved Orion MR into two components separated by $\sim0\rlap.{''}3$ (see Figure 3), which correspond to 116 AU
at a distance of 388 pc (Kounkel et al. 2017). If we assume that these two sources form a binary system and that the orbit is circular and in the plane of the sky, we estimate that the total mass associated with Orion MR
is 6.5 $M_\odot$. This result requires confirmation in the future. For example, if we remove from the fit the first epoch, we get a slope of 0.40$\pm$0.12 degrees yr$^{-1}$, significant only to 3.3-$\sigma$. On the other hand, the luminosity of
2,000 $L_\odot$ determined by Greenhill et al. (2004) implies a star of 8 $M_\odot$, using the mass-luminosity relationship of Wang \& Zhong (2018). This approximate agreement supports the reality of the rotation.

\begin{figure}
\centering
\vspace{-2.8cm}
\includegraphics[angle=0,scale=0.7]{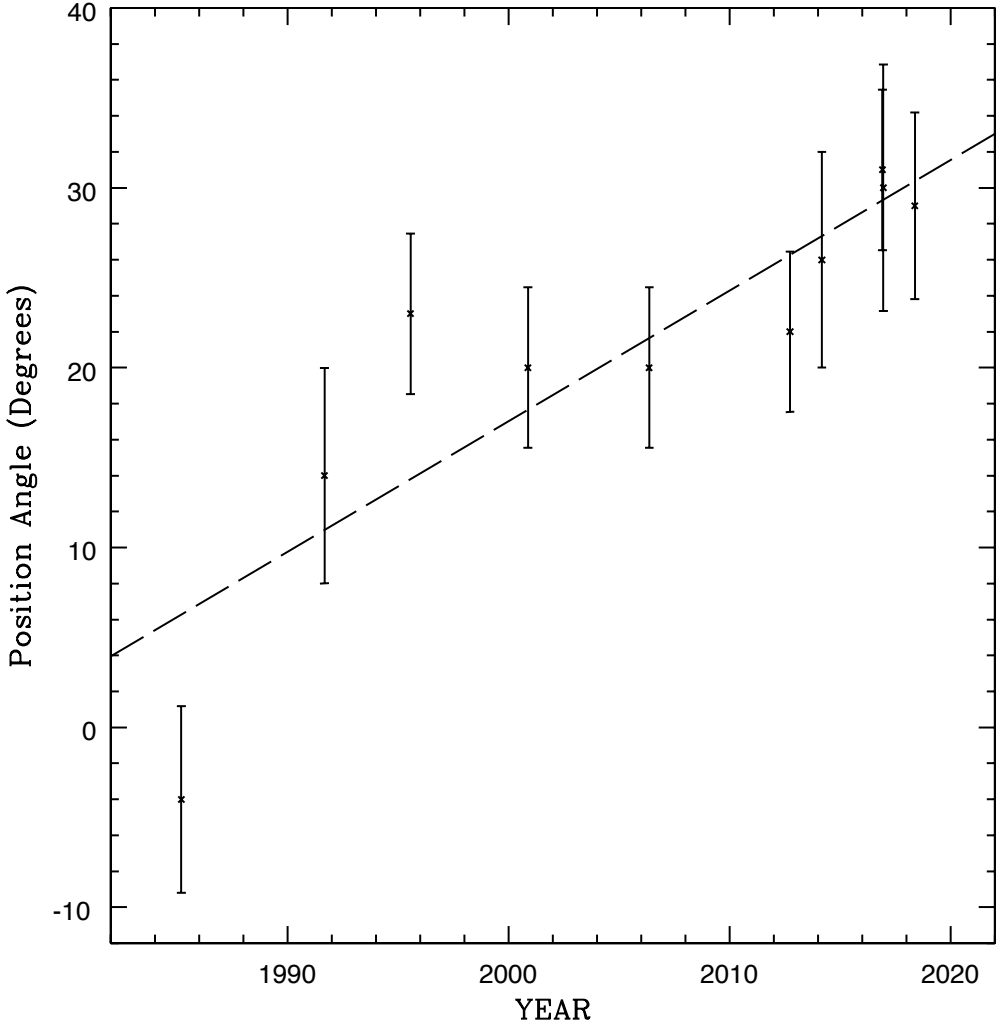}
\vskip-2.0cm
\caption{\small Deconvolved position angle for Orion MR as a function of time. The dashed line is a least-squares linear fit to the data.}
\label{fig4}
\end{figure}
\pagebreak

\section{The proper motions of IRc23 and Zapata 11}

Dzib et al. (2017) found that the radio sources IRc23 and Zapata 11 also seem to show proper motions receding from the same point 
from which the other sources are escaping. We use the four new epochs to test this hypothesis and fully confirm the significant large proper motions of these
two additional sources. In Figures 5 and 6 we show their positions as a function of time and in Table 4 we summarize the proper motions of the
six sources that are moving away from the central point of the explosion (BN, I, Orion MR, x,  IRc23 and Zapata 11). Finally, in Figure 7 we show an image with the proper motions of the six sources reported to have large proper motions.
The proper motions are consistent with a common epoch of ejection for all sources (about 500 years ago), except for the case of IRc23, that appears to have been ejected only $\sim$300 years ago.

It is interesting to compare these recent ejection phenomena with the historic ejection events associated with the ONC. The classic runaway pair is formed by the OB stars
$\mu$ Col and AE Aur, that were ejected in opposite directions with velocities of $\sim$100 km s$^{-1}$ from the vicinity of the ONC about 2.6 million years ago (Blaauw \& Morgan 1954).
Hoogerwerf et al. (2001) find that the massive highly-eccentric double-lined spectroscopic binary $\iota$ Ori has proper motions that imply that it was very close to $\mu$ Col and AE Aur 2.6 million years ago.
These authors propose as an explanation for this ejection of three stellar systems a binary-binary encounter. In the case of the recent ejection of multiple stellar systems in Orion BN/KL
it could be of interest to explore numerically the interaction of a binary with a compact cluster. Finally, the $\beta$ Cephei variable 53 Ari has a peculiar velocity of 48 km s$^{-1}$ relative to its neighbors (Tetzlaff et al. 2011) and 
appears to have been ejected from the ONC some 4-5 million tears ago. Then, in these historical events we seem to have different ejection epochs, as in the case of IRc23.

\begin{figure}
\centering
\vspace{-0.0cm}
\includegraphics[angle=0,scale=0.65]{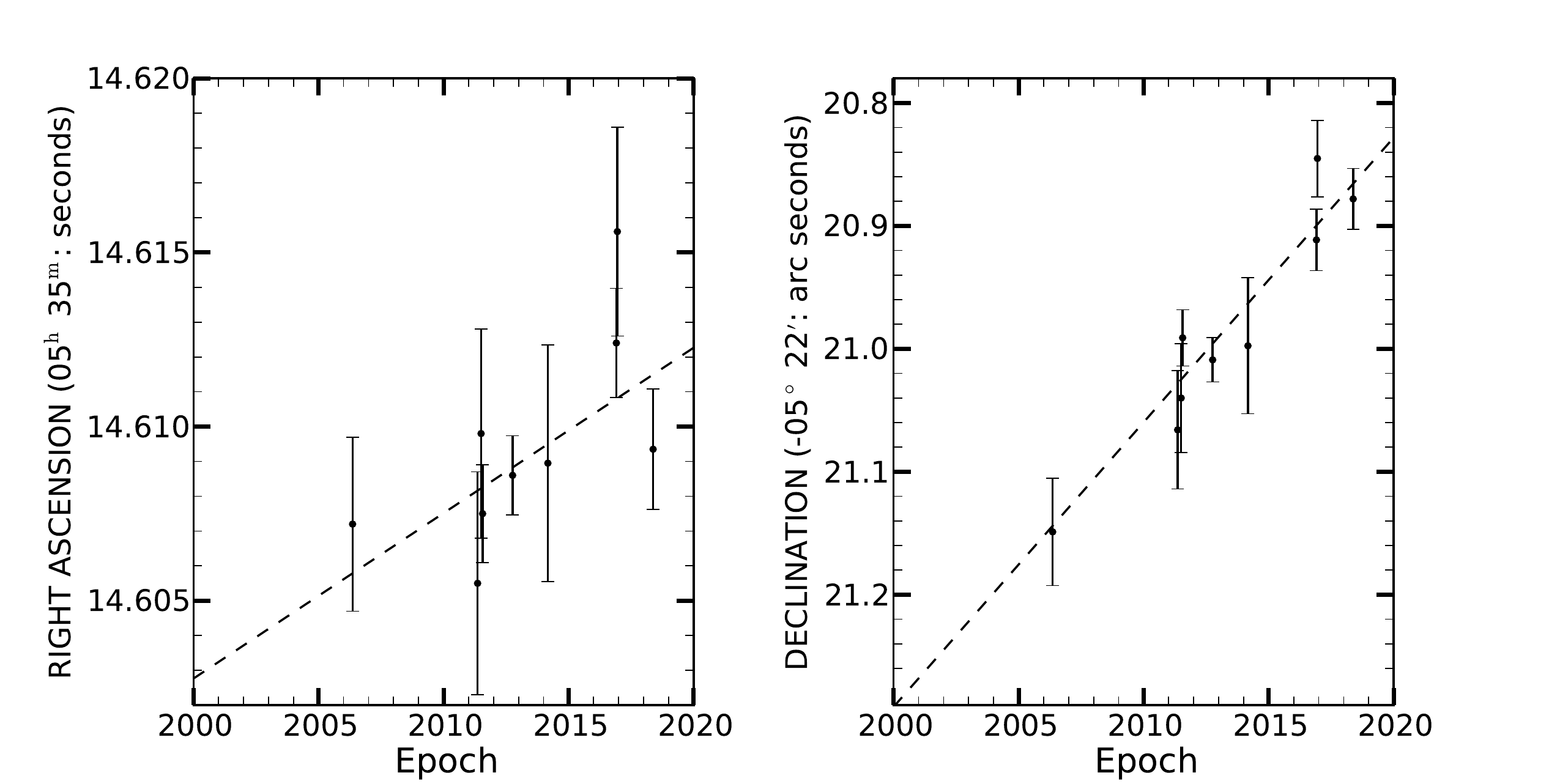}
\vskip-0.0cm
\caption{\small Right ascension (left) and declination (right) of the source IRc23 as a function of time. The error bars include a systematic component
added in quadrature to make $\chi^2$ = 1. The dashed lines are least-squares linear fits to the data. The parameters of the fits are given in the text. 
}
\label{fig5}
\end{figure}
\pagebreak

\begin{figure}
\centering
\vspace{-0.0cm}
\includegraphics[angle=0,scale=0.65]{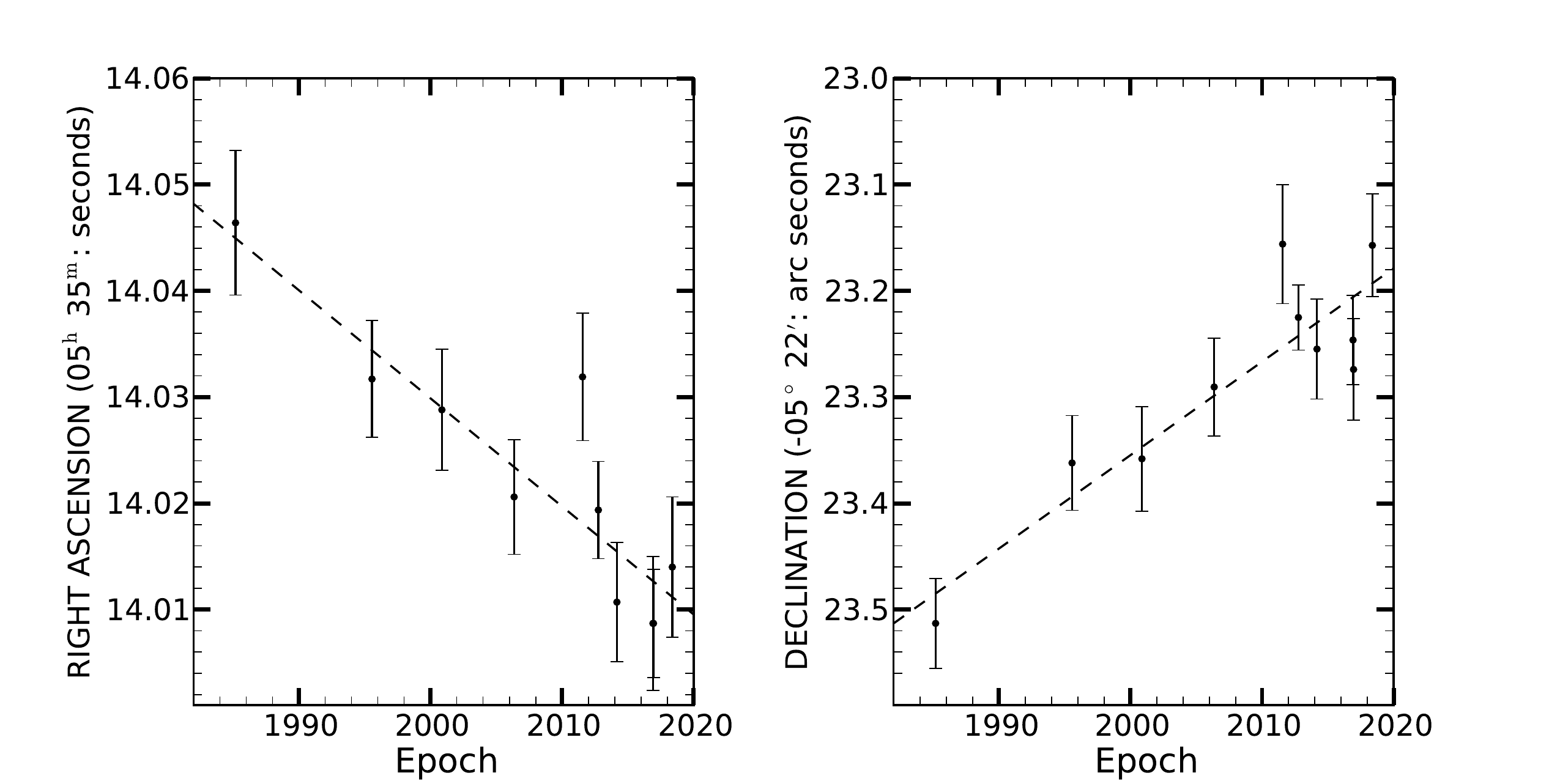}
\vskip-0.0cm
\caption{\small Right ascension (left) and declination (right) of the source Zapata 11 as a function of time. The error bars include a systematic component
added in quadrature to make $\chi^2$ = 1. The dashed lines are least-squares linear fits to the data. The parameters of the fits are given in the text. 
}
\label{fig6}
\end{figure}
\pagebreak

\begin{figure}
\centering
\vspace{-0.0cm}
\includegraphics[angle=0,scale=0.7]{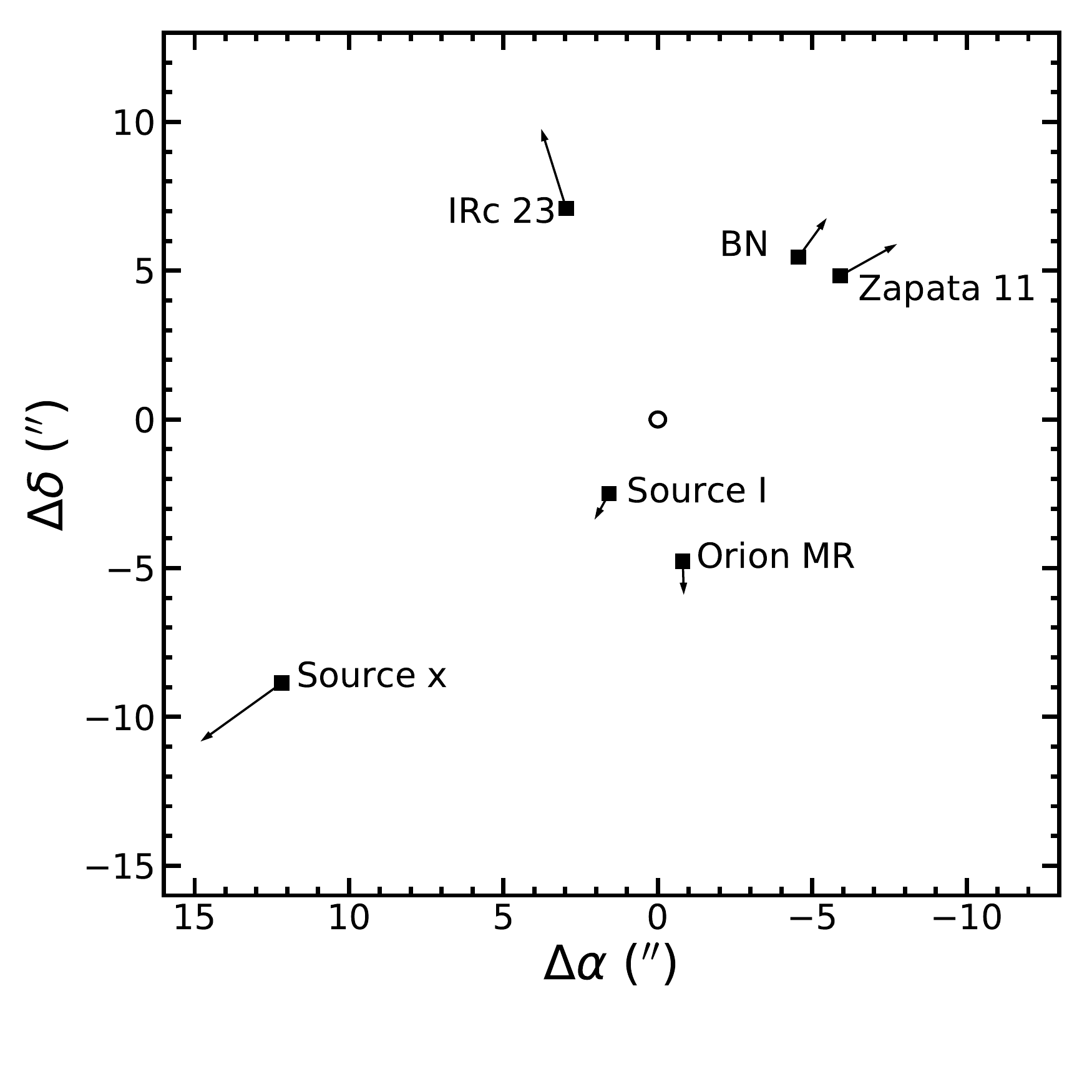}
\vskip-0.0cm
\caption{\small Positions and proper motions of the six sources reported with large proper motions. The size of the arrows indicate the proper motions for a
period of 100 years. The data are taken from Rodr\'\i guez et al. (2017), Luhman et al. (2017) and this paper.
The empty circle marks the position of the center of the explosion as determined by Rodr\'\i guez et al. (2017), $\alpha(2000) = 05^h~ 35^m~ 14\rlap.^s41; $
$\delta(2000) = -05^\circ~ 22'~ 28\rlap.{''}2.$
}
\label{fig7}
\end{figure}
\pagebreak

\begin{deluxetable}{lccccc}
\tabletypesize{\scriptsize}
% \tablewidth{18.5cm}
\tablecaption{Proper Motions of the Six Sources in Orion BN/KL}
%\small
\tablehead{
\colhead{}  & \multicolumn{4}{c}{Proper Motions$^a$} 
& \colhead{} \\
\colhead{Source} & \colhead{$\mu_\alpha cos \delta$} & \colhead{$\mu_{\delta}$}
& $\mu_{total}$ & $PA$ & \colhead{Reference} }
\startdata
BN &  $-7.0\pm0.4$ &  $10.0\pm0.4$ & $12.2\pm0.4$ & $-35^\circ\pm2^\circ$ & Rodr\'\i guez et al. (2017) \\
I &  $2.9\pm0.4$ &  $-5.4\pm0.4$ & $6.1\pm0.4$ & $152^\circ\pm4^\circ$ & Rodr\'\i guez et al. (2017) \\
Orion MR &  $0.2\pm0.7$ &  $-7.5\pm0.7$ & $7.5\pm0.7$ & $178^\circ\pm5^\circ$ & This paper\\
x &  $23.3\pm1.4$ &  $-17.4\pm1.4$ & $29.1\pm1.4$ & $127^\circ\pm3^\circ$ & Luhman et al. (2017) \\
IRc23 &  $7.1\pm3.0$ &  $23.2\pm3.1$ & $24.3\pm3.1$ & $17^\circ\pm7^\circ$ &  This paper \\
Zapata 11 &  $-15.2\pm2.8$ &  $8.8\pm1.3$ & $17.6\pm2.5$ & $-60^\circ\pm8^\circ$ & This paper \\
\enddata
\tablenotetext{a}{The proper motions are given in units of $mas~yr^{-1}$. The position angle (PA) is given in degrees.}
%\tablenotetext{b}{Major axis$\times$minor axis in arcsec; PA in degrees. The deconvolved dimensions of the source are corrected for the effect of bandwidth smearing.
%In all cases the minor axis of the source remains unresolved.}
%% You can append references to a table using the \tablerefs command.
\end{deluxetable}

\section{Conclusions}

Our main conclusions can be summarized as follows.

1. While the infrared source n has been found by several groups (Luhman et al. 2017; Kim et al. 2019) to be stationary, its proposed VLA radio counterpart has a clear proper motion to the
south. We conclude that the explanation to this discrepancy is simple: the two sources are independent. Source n is a single source that is approximately
stationary, while the radio source Orion MR is a doble source (presumably a binary system) that is moving at $\sim$14$\pm$2 km s$^{-1}$ with respect to the Orion reference frame.

2. Using the derived proper motions we find that, by year 1475, Orion MR was at a position consistent with the centroid of the ejection of gas and stars,
as determined by Rodr\'\i guez et al. (2017). Therefore this source forms part of the remarkable explosive event that took place in the
Orion BN/KL region.

3. The major axis of Orion MR seems to show a rotation with a period of 490$\pm$100 years. Assuming that we have a binary system in a circular orbit
on the plane of the sky, we conclude that the total mass of the system is 6.5 $M_\odot$.

4. We used the new data to confirm the proper motions of IRc23 and Zapata 11. The latter source appears to be part of the explosive event, while IRc23 may have been ejected only 300 years ago. Up to now, there are six compact sources
that recede from the central position of the remarkable explosive phenomenon present in Orion BN/KL, five of them consistent with an ejection epoch of 1475. The relatively large number of sources rules out as a possible mechanism the classic three-body 
scenario since then only two escaping bodies are expected: a tight binary plus the third star involved in the encounter. However, more complex interactions such as that of a massive binary with a compact
cluster may explain the BN/KL phenomena.

\acknowledgements
% We thank an anonymous referee for a careful revision of our paper that improved its clarity.
LFR, SL, LL and LZ acknowledge the financial support of
PAPIIT-UNAM IN105617, IN101418, N110618 and IN112417 and CONACyT  238631 and 280775. % and of CONACyT (M\'exico).
This research has made use of the SIMBAD database,
operated at CDS, Strasbourg, France.

\facility{VLA}

\software{AIPS van Moorsel et al. (1996); CASA McMullin et al.(2007)}

\end{document}